\documentclass[twocolumn,showpacs,preprintnumbers,amsmath,amssymb,pra]{revtex4}

\usepackage{graphics}
\usepackage{epsfig}
\usepackage{bm}

\def\pop{Phys.\ Plasmas\ }
\def\etal{{\em et al.}}

\def\beq{\begin{equation}}
\def\eeq{\end{equation}}
\def\bec{\begin{center}}
\def\eec{\end{center}}

\def\reff#1{(\ref{#1})}

\def\subsc#1{{\mbox{\rm\scriptsize #1}}}

\def\Etot{E_\mathrm{tot}}
\def\Eig{E_\mathrm{ig}}
\def\Qb{Q_\mathrm{b}}

\def\Wcmcm{\mbox{\rm Wcm$^{-2}$}}

\def\N3d{N_\subsc{3D}}

\def\Rinit{R_0}

\def\XeN{\mathrm{Xe}_N}

\def\omegaMie{\omega_\mathrm{Mie}}

\def\omegalaser{\omega_\mathrm{l}}
\def\lambdalaser{\lambda_\mathrm{l}}

\def\laserE{E_\mathrm{l}}
\def\vectlaserE{\vect{E}_\mathrm{l}}
\def\Ionp{I_\mathrm{p}}
\def\chZ{\mathcal{Z}}
\def\chZmax{\chZ_\mathrm{max}}
\def\chZmin{\chZ_\mathrm{min}}
\def\chZav{\chZ_\mathrm{av}}

\def\vekt#1{\bm{#1}}
\def\vect#1{\vekt{#1}}

\def\vektR{\vekt{R}}

\def\vektR{\vekt{R}}

\def\vektE{\vekt{E}}

\def\vektEsc{\vektE_\mathrm{sc}}
\def\Eig{E_\mathrm{ig}}
\def\Escx{E_\mathrm{sc}^x}

\def\Edach{E_0}

\def\Ehat{\Edach}

\def\diff{\,\mbox{\rm d}}

\begin{document}
\title{Optimizing the ionization and energy absorption of laser-irradiated clusters
}
\date{\today}
\author{M.\ Kundu}
\author{D.\ Bauer}
\affiliation{Max-Planck-Institut f\"ur Kernphysik, Postfach 103980,
69029 Heidelberg, Germany}
\date{\today}

\begin{abstract}
It is known that rare-gas or metal clusters absorb incident laser energy very efficiently.
However, due to the intricate dependencies on all the  laser and cluster parameters it is difficult to predict under which circumstances ionization and energy absorption is optimal.
With the help of three-dimensional particle-in-cell simulations of xenon clusters (up to 17256 atoms)   we find that for a given laser pulse energy  and cluster an optimum wavelength exists which
corresponds to the approximate wavelength of the transient, linear Mie-resonance of the ionizing cluster at an early stage of negligible expansion. %
In a single ultrashort laser pulse, the linear resonance at this optimum wavelength yields much higher absorption
efficiency than in the conventional, dual-pulse pump-probe set-up of linear resonance during cluster expansion. 
\end{abstract}


\maketitle

\section{Introduction}\label{sec1Ch5}
The interaction of rare-gas and metal clusters with intense laser light has drawn close attention during the last ten years. Reasons for this large interest are (among others) the  high charge states  
\cite{ditmNature,ditm97,ditm97a,ditm97b,kum01a,kum01,kris04,kris06jpb,lezius98}  and the high energies of both ions \cite{ditmNature,ditm97,ditm97a,ditm97b,kum01,kris04,kris06jpb,lezius98,kum02,spring00,spring03}
and electrons \cite{kum02,shao96,kum03,spring03,chen02pop} observed (see \cite{saal06} for a recent review).

The laser-cluster interaction scenario may be qualitatively summarized as follows. Electrons leave their ``parent'' atoms (``inner ionization'') and absorb further laser energy while moving in the cluster potential formed by the ionic background.  Some electrons may leave the 
cluster (``outer ionization''), leaving behind a net positively charged nanoplasma.   The total electric field consists of the laser plus the space charge field. It may exceed the pure laser field in certain spatial regions and thus may liberate further electrons from their parent ions (which would remain bound if there was the laser field alone). This enhanced inner ionization 
is called ``ionization ignition'' \cite{rose97, bauer03, bauer04, ishi00, megi03, crist04}. 
Ionization ignition locally continues until the total field drops below the threshold field required to liberate further electrons.  
The ionic background expands because of Coulomb repulsion and hydrodynamic pressure, which ultimately leads to the energetic ions observed in experiments.
Because of the complex interplay between inner ionization, outer ionization, and ionic expansion it is far from trivial to predict quantitatively mean or highest charge states, the absorbed laser energy, or other observables and their dependence on the laser parameters and the cluster decomposition.

One of the goals in laser-cluster experiments and simulations 
is to convert as much laser energy as possible into energetic particles. This can be achieved by optimizing the outer ionization degree, i.e., by removing as many electrons as possible from the cluster in order to generate high charge states so that the asymptotic ion energy (and thus the total absorbed energy) after Coulomb explosion is largest. 


One way to increase the charge states and the ion energy 
is to dope a cluster with atomic/molecular species of low ionization 
potential \cite{lastX1,hohen05}.  
An almost two-fold increase of the highest charge states 
were obtained experimentally with
argon clusters doped with water molecules \cite{jhaX1}. 

Experimental results for xenon and silver clusters 
embedded in helium droplets were reported in Ref.~\cite{dop05}. 
The pulse duration and the sign of the
chirp of a laser-pulse also affect ion charge states and ion energies \cite{fukuda03}.
Enhanced inner ionization of rare-gas and metal clusters irradiated by 
a sequence of dual laser pulses 
were observed \cite{spring00,dop05,doeppner06,fennel07} experimentally. 
In these kinds of experiments one should adjust the delay time between pump and probe pulse such that  
the cluster expands sufficiently to meet the linear resonance $\omegaMie=\omegalaser$
with the probe pulse, where  $\omegaMie$ is the Mie-plasmon frequency and $\omegalaser$ is the laser frequency.
Vlasov simulations \cite{dop05} and semi-classical simulations \cite{sied05}
of a small $\mathrm{Xe_{40}}$ cluster subject to such a pump-probe setup
 showed an enhancement of the ion charge states. 
An optimum control multi-pulse simulation has also been performed \cite{mart05}.

In this work we investigate 
the effect of the laser wavelength by three-dimensional 
particle-in-cell (PIC) simulations. The goal is to find an optimum wavelength for a fixed laser intensity and a given cluster. 
At this optimum wavelength (which turns out to be in the ultraviolet (UV) regime for the Xe clusters under consideration) a single ultrashort laser pulse is shown to be much more 
efficient than the ``conventional'' dual-pulse pump-probe setup.

Experimental signatures of enhanced x-ray yields and high charge
states at short wavelengths \cite{rhodes94,rhodes98,rhodes97} indicate a clear impact of the laser wavelength
on the laser-cluster interaction. Free electron laser (FEL) cluster experiments \cite{wab02} at the DESY facility, Hamburg,
and a recent x-ray laser-cluster experiment \cite{namba07} down to wavelengths $<100$~nm 
also showed enhanced ionization.

Contrary to our findings recent molecular dynamics simulations \cite{pet06} concluded that (i) there is no influence of the laser 
wavelength on the charging of clusters in the regime 
$100-800$~nm for a laser intensity $\approx 10^{16}\Wcmcm$ and (ii) that linear resonance plays no role, thus threatening the basis of the nanoplasma model \cite{ditm96}. Similar conclusions were reported by the same authors in Refs.~\cite{pet07,pet05}.

We consider short laser pulses in this work. Most of the earlier works were 
reported for the long-pulse regime where linear resonance (LR) absorption
\cite{ditm96,dop05,koell99,zam04,last99,saal03,fenn04,sied05,mart05} 
occurs during the 
expanding phase of a cluster when the Mie-plasma frequency drops sufficiently so that the laser frequency can be met. 
At the time of LR the space charge field inside the cluster is strongly
enhanced, leading to efficient ionization ignition.

The paper is organized as follows.
In Sec.~\ref{sec2Ch5} we briefly describe the simulation method and discuss  the ionization of a cluster by a short laser pulse in Sec.~\ref{sec3Ch5}.
In Sec.~\ref{sec4Ch5} pump-probe simulation results are presented while
Sec.~\ref{sec5Ch5} is devoted to the laser wavelength dependence of the cluster dynamics. A possibility to achieve $100\%$ outer ionization is also discussed in Sec.~\ref{sec5Ch5} before we summarize the work in Sec.~\ref{sec6Ch5}.
Unless stated otherwise we use atomic units (a.u.).

\section{Details of the simulation}\label{sec2Ch5}
Details of our PIC code are already described in 
Refs.~\cite{kundu06b,kundu07}.

For the inner ionization we apply the so-called Bethe-rule or 
over-the-barrier ionization (OBI) model \cite{bethe}. According to OBI
an atom or ion is ionized if the total field satisfies 
\beq
\vert \vectlaserE(t) + \vektEsc(\vektR_j,t) \vert\ge\Ionp^2(\chZ)/4\chZ
\label{bethe1Ch4}\eeq at individual ion locations $\vektR_j$.
Here $\chZ$ is the charge state (after ionization), $\Ionp(\chZ)$ is the respective ionization potential and
$\vektEsc(\vektR_j,t)$ is the space charge field.
In the absence of $\vektEsc(\vektR_j,t)$ ionization is caused
by the laser field, known as optical field ionization (OFI).

In this work we shall vary the wavelength down to $100$~nm at an intensity $5\times10^{16}\Wcmcm$, which raises the questions
(i) whether such lasers are available and  
(ii) whether the Bethe-rule~\reff{bethe1Ch4} is applicable. 
With the development of new generation FEL lasers \cite{and00,ayv02} the answer to (i) is clearly affirmative.
As regards the ionization model (ii), at short wavelengths ionization  rather proceeds via multiphoton ionization than via tunneling or over-the-barrier ionization so that the Bethe-model (where the ionization probability switches from zero to unity once a certain threshold field is reached) may not yield the precise charging dynamics of the clusters at short wavelengths. However, the final charge state distribution should remain unaffected by the details of the ionization model \cite{pet05} at least qualitatively.

Only collisionless absorption mechanisms are incorporated in standard PIC simulations. 
The neglect of particle collisions is an approximation which is the more valid the smaller the clusters, the higher the laser intensities, and the longer the laser wavelengths are (see, e.g., \cite{ishi00,megi03,crist04,bauer04,sied04,crist05}). Since collisions can be expected to increase both energy absorption and ionization, our results from collisionless PIC calculations may be considered close to reality for wavelengths $\gtrapprox 400$\,nm but slightly underestimating  the real charge states and the real absorbed energy for shorter wavelengths.

\section{Ionization of a cluster by a single short pulse}\label{sec3Ch5}
First we study the response of a xenon cluster 
in a linearly polarized 
$n=8$-cycle laser pulse of electric
field strength $\laserE(t)=\Ehat\sin^2(\omegalaser t /2n)\cos(\omegalaser t)$ 
and wavelength $\lambdalaser = 800$~nm.
Different ionic charge states are self-consistently produced  during the laser pulses according to the Bethe rule~\reff{bethe1Ch4}.

%
\begin{figure}
\includegraphics[width=0.45\textwidth]{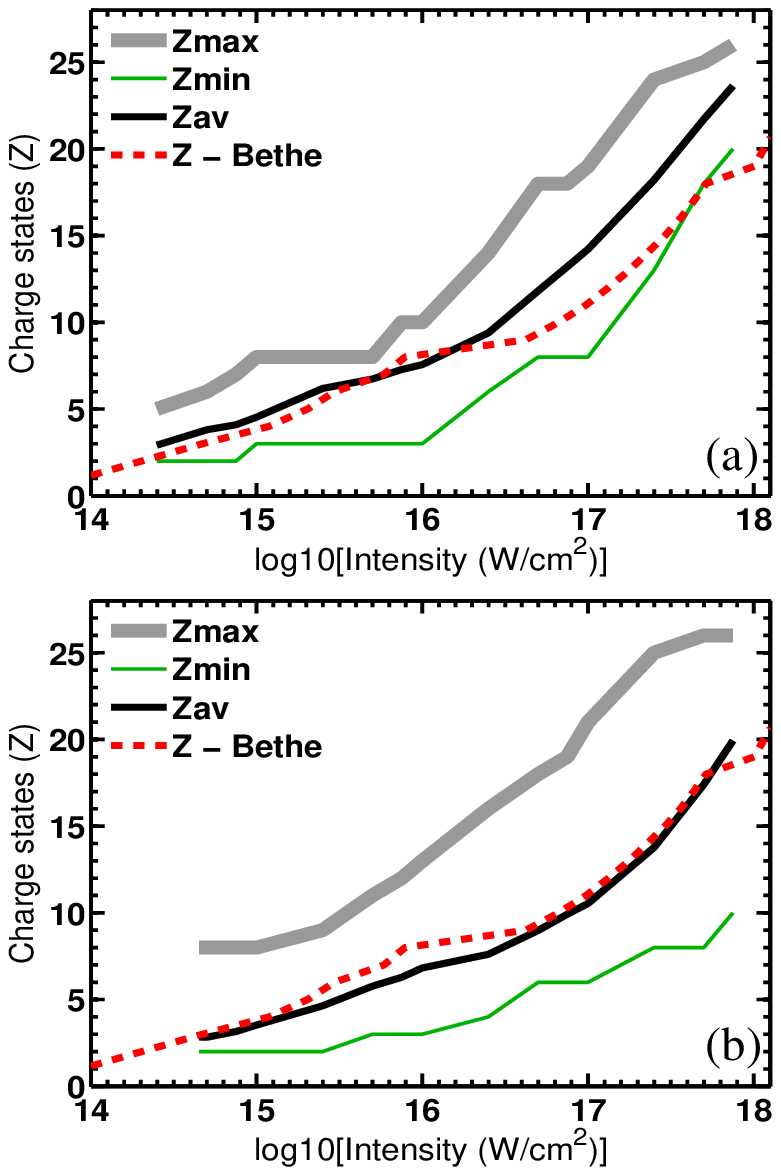}
\caption{(Color online) 
Maximum ion charge $\chZmax$ (thick solid),
minimum ion charge $\chZmin$ (thin solid),
average ion charge $\chZav$ (bold black) and the ion charge
predicted by OFI alone (dashed)
vs peak laser intensity for 
(a) a $\mathrm{Xe_{2176}}$ cluster of radius $\Rinit \approx 3.54$~nm
and (b) a $\mathrm{Xe_{17256}}$ cluster of radius $\Rinit \approx 7$~nm
in an $n=8$ cycle laser pulse
$\laserE(t)=\Ehat\sin^2(\omegalaser t/2n)\cos(\omegalaser t)$ of wavelength 
$\lambdalaser = 800$~nm. 
\label{Fig1Ch5}
}
\end{figure}

Figure~\ref{Fig1Ch5}a shows the
maximum charge state $\chZmax$, 
the minimum charge state $\chZmin$, 
and the average charge state $\chZav$ 
(defined as the total charge of the cluster
divided by the number of atoms $N$) and the charge state predicted by
the OFI (``Z-Bethe'' curve) vs peak laser intensity 
for a $\XeN$ cluster ($N = 2176$) of 
initial radius $\Rinit \approx 3.54$~nm after the pulse (i.e., after $\approx22$~fs).
The maximum charge state $\chZmax$ varies from  $\chZ =5$ to $\chZ = 26$ 
as the laser intensity increases from 
$2.5\times 10^{14}\Wcmcm$ to $7.5\times 10^{17}\Wcmcm$. The higher value of $\chZmax$ above
the value predicted by the OFI is clearly
due to ionization ignition.
Those maximum charge states $\chZmax$ are mainly 
acquired by the ions at the 
cluster periphery where the space charge field is highest.
Inside the cluster the total field falls below the ionization 
thresholds due to the decreasing space charge produced by the ionic background as well as due to the screening of the 
laser field by the cluster electrons. 
The ions close to the cluster center have minimum charge states
$\chZmin = 2-20$ at laser intensities
between $2.5\times 10^{14}\Wcmcm - 7.5\times 10^{17}\Wcmcm$.
The value of $\chZmin$ remains much lower than predicted by the 
OFI for almost all laser intensities $<5.0\times 10^{17}\Wcmcm$. 
The average charge $\chZav$ remains close to (but slightly  higher than) the OFI predicted values 
at intensities $<7.5\times10^{15}\Wcmcm$. Also
$\chZmax = 8$ and $\chZmin = 3$ do not change between the 
intensities $10^{15}\Wcmcm - 7.5\times 10^{15}\Wcmcm$ but $\chZav$ increases slowly 
as more ions from the cluster center towards the periphery acquire higher charge states
$3\rightarrow 8$. 
The value of $\chZmax$ remains constant, $\chZ = 8$, due to the removal of 
all electrons from the $5s^2p^6$ shell of the Xe atoms close to the cluster boundary. 
As the intensity $\approx 7.5\times10^{15}\Wcmcm$ is approached the 
laser field is strongly shielded from the central part of the cluster, 
and outer ionization as well as 
ionization ignition tend to saturate. As a consequence $\chZav$ grows slowly
between the intensities $\approx 5\times10^{15}\Wcmcm - 10^{16}\Wcmcm$. 
Unless a threshold intensity $\approx 10^{16}\Wcmcm$  
is crossed further electrons
from the cluster cannot be removed, which was already seen in  
previous model and numerical calculations \cite{kundu06a,kundu06b,mulser05,tagu04}.
At higher intensities $>10^{16}\Wcmcm$  outer ionization and ionization ignition increases again, leading to an increase of
$\chZav$ beyond the values predicted by the OFI due to the strong increase of both $\chZmax$ and $\chZmin$.

It is commonly believed that ionization ignition becomes increasingly pronounced with increasing cluster size.   
Figure~\ref{Fig1Ch5}b shows $\chZmax$, $\chZmin$, $\chZav$, 
and the charge states predicted by
the OFI vs the peak laser intensities
for a bigger $\XeN$ cluster ($N = 17256$) 
of initial radius $\Rinit \approx 7$~nm. $\chZmax$ varies between $8-26$, exceeding again the
charge states predicted by OFI alone.
Below the intensity $10^{17}\Wcmcm$ $\chZmax$ is higher by a factor of $\approx 2$ 
compared to the OFI value (``Z-Bethe'' curve).
Although $\chZmax$ remains 
much higher, the average ion charge $\chZav$ (in Fig.~\ref{Fig1Ch5}b) is below the 
charge states predicted by the OFI for most of the laser intensities.
Most of the ions closer to the cluster center  
acquire charge states $\chZmin = 2-10$ which are even lower than for
the smaller cluster (Fig.~\ref{Fig1Ch5}a) at the corresponding intensities.
Hence, ionization ignition is indeed 
responsible for the highest charge states $\chZmax$ which increase with the cluster size (as seen in Fig.~\ref{Fig1Ch5}). However, exactly because of the same mechanism  a bigger cluster will capture more electrons (whose outer ionization 
 would require much higher laser intensities than in the case of a smaller cluster). 
The presence of more electrons in the central region 
will screen the laser field more efficiently. 
As a result both $\chZav$ as well as $\chZmin$ 
(in Fig.~\ref{Fig1Ch5}b) drop
below the corresponding values for the smaller cluster (Fig.~\ref{Fig1Ch5}a).

We conclude that an increasing cluster size (and thus increased ionization ignition of, at least, the ions located close to the cluster boundary) does not always lead to a higher average charge state.
Our aim is to increase not only the highest charge states but also 
the average ion charge beyond the OFI predicted value
 through the charging of more ions in the central part of the cluster.
In the following sections we study several approaches to achieve this goal.

\section{Ionization by delayed pulses: a pump-probe simulation}
\label{sec4Ch5}
In this section we illustrate 
the ``pump-probe'' method frequently employed in laser-cluster experiments. 
In this method an initial pump-pulse 
ionizes the cluster. The cluster expands freely before, after a delay time, a probe-pulse hits the expanding cluster. The interaction of this probe pulse with the cluster will sensitively depend on the cluster size and thus on the delay time.
We revisit such a scenario in our current work since it 
will allow us to compare the efficiency of laser energy absorption
for such a standard pump-probe method with the single UV pulse scenario which will be introduced in Sec.~\ref{sec5Ch5}.

The laser field profile is of the form 
$\laserE(t)=\Ehat\sin^2(\pi t/n T)\cos(\omegalaser t)$ for both 
pump and probe pulse. 
The time period $T$ is chosen with respect to the wavelength
$800$~nm. For, say, $n=4$ the product $n T$ determines the total pulse duration $\approx 11$~fs. 
The pulse envelope and intensity are kept the same
for all cases under study, i.e., the laser energy in all pulses is the same too. 

\begin{figure}
\includegraphics[width=0.45\textwidth]{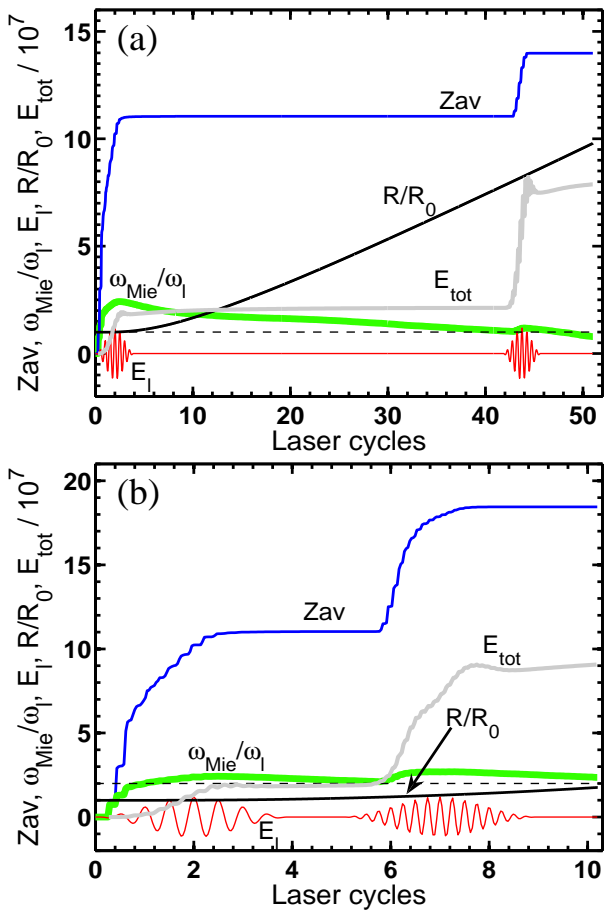}
\caption{(Color online)
Average ion charge $\chZav$, scaled Mie-frequency $\omegaMie/\omegalaser$, 
laser field $\laserE$ (in atomic units), normalized cluster expansion radius $R/\Rinit$ and total
absorbed energy $\Etot$ (in atomic units) vs time (in 800~nm laser cycles) for a 
$\mathrm{Xe_{17256}}$ cluster of radius $\Rinit \approx 7$~nm. 
The peak intensity $5\times10^{16}\Wcmcm$ is the same for 
(a) pump of wavelength $400$~nm (probe, 400~nm) and 
(b) pump of wavelength $400$~nm (probe, 200~nm). 
The laser field is of the form
$\laserE(t)=\Ehat\sin^2(\pi t/nT)\cos(\omegalaser t)$ with  
$n=4$ and one laser cycle $T$ corresponding to the wavelength $800$~nm.
A minor increase in $\Etot$ after the pulses is an artifact of PIC simulations.
\label{Fig2Ch5}
}
\end{figure}

Figure~\ref{Fig2Ch5}a shows the results for the
$\mathrm{Xe_{17256}}$ cluster of initial radius $R_0\approx7$~nm 
at an intensity $5\times 10^{16}\Wcmcm$
when both the pump and the probe pulse  have the same wavelength 400~nm. 
The average charge $\chZav$, the scaled Mie-frequency $\omegaMie/\omegalaser$, the total absorbed energy 
$\Etot$ (electrostatic field energy plus the kinetic energy 
of electrons and ions), the normalized cluster radius $R(t)/R_0$, and the laser fields are plotted vs time  (in units of the period $T$).
During the first four laser cycles of the pump-pulse
the average charge state rises to $\chZav \approx 11$,
the frequency $\omegaMie/\omegalaser \approx 2.5$ and 
$\Etot \approx 2.0\times 10^{7}$ while the cluster expansion is insignificant.  
The total energy $\Etot \approx 2.0\times 10^{7}$ corresponds to the 
average energy absorbed per ion $\Etot/N \approx 31.4$~keV.  
After the pump-pulse the cluster evolves freely and $\chZav$, $\Etot$ remain 
unchanged but $\omegaMie/\omegalaser$ drops due to the expansion.
Note that the cluster radius $R(t)$ (defining the cluster boundary)  corresponds to the distance of the most
energetic ions from the cluster center. At the boundary, however, the cluster potential is anharmonic.  Hence using $R(t)$ for the calculation of the Mie-frequency $\omegaMie(t) = \sqrt{N \chZav / R^3(t)}$ the latter is underestimated. 
Instead we use the definition 
$\omegaMie(t) = \sqrt{\Qb(t) / R_0^3}$ (as in Ref.~\cite{kundu07}) where $\Qb(t)$ is the total 
ionic charge within the {\em initial} cluster radius $R_0$ where the cluster potential is close to harmonic at all times. 

After $44$ laser cycles $\omegaMie$ approaches the linear resonance (dashed line) with respect to the fundamental 400~nm, i.e., $\omegaMie/\omegalaser = 1$.  
The probe pulse of wavelength 400~nm
is applied with a delay of $\approx 42$ laser cycles
such that the peak of the pulse approximately coincides with the 
 resonance time.
Due to the linear resonance the  
average charge and the absorbed energy rises abruptly 
up to the value $\chZav = 14$
and $\Etot \approx 7.5\times 10^{7}$, respectively.
Such a pump-probe simulation clearly illustrates that the linear 
resonance indeed plays a role in the cluster dynamics. More energy is absorbed, leading to higher charge states. 
These results are in agreement
with hydrodynamic and Vlasov simulations\cite{spring00,dop05}. 
However, linear resonance is met only after a relatively long time when the 
cluster has already expanded significantly (as seen $R(t)/R_0 \approx 8$ 
in Fig.~\ref{Fig2Ch5}a). Ionization ignition and laser energy absorption in such a 
low density plasma is expected to be less efficient compared to the case where
linear resonance occurs {\em before} the cluster expands significantly.

While keeping the 400~nm pump as above we now assume a probe wavelength of 200~nm for the purpose of hitting the linear resonance at an earlier time when the cluster is more compact. The energy in the probe pulse is the same as in Fig.~\ref{Fig2Ch5}a.
Figure~\ref{Fig2Ch5}b shows the result analogous to Fig.~\ref{Fig2Ch5}a. 
The average charge and the 
absorbed energy now increase up to $\chZav = 18.5$ and 
$\Etot \approx 9\times 10^{7}$ which are
 higher than in Fig.~\ref{Fig2Ch5}a after the probe. 
With the pulse energies being the same in both cases a higher efficiency of energy absorption in the second scheme (Fig.~\ref{Fig2Ch5}b) is obvious.
The reason is the smaller cluster size at the time of linear resonance ($R(t)/\Rinit< 1.5$) and the higher space charge field related to it.
Similar findings from experiments have been reported in Ref.~\cite{spring00}.
In passing we note that the average charge $\chZav \approx 11$ in 
Fig.~\ref{Fig2Ch5} due to the pump (at 400~nm) exceeds $\chZav\approx 8$ in Fig.~\ref{Fig1Ch5}b (at 800~nm) for the same 
cluster and the same laser intensity $5.0\times10^{16}\Wcmcm$ despite
the higher pulse-energy in Fig.~\ref{Fig1Ch5}b because of the twice longer pulse. 

In the following section we study the wavelength dependence of the average charge states and the laser energy
absorption.

\section{Ionization at different wavelengths}\label{sec5Ch5}
Does the average charge state and the absorbed energy for a given cluster increases with decreasing laser wavelength?  One could expect that for a certain wavelength the linear resonance during the initial ionization stage when the Mie-frequency rises from zero to its maximum value becomes important. For long wavelengths this early resonance is passed so quickly due to the rapid charging of the cluster that any indication of a resonance is washed out.

We assume the same laser field profile 
$\laserE(t)=\Ehat\sin^2(\pi t/n T)\cos(\omegalaser t)$ 
as in Sec.~\ref{sec3Ch5} with the same pulse duration, pulse energy, 
and intensity $5.0\times10^{16}\Wcmcm$ so that the number of laser cycles in the pulse depends on the wavelength.
The laser wavelength is varied  in the range $800-100$~nm. Note that in the following we specify times and pulse durations in units of laser periods at 800\,nm (corresponding to $T \approx 2.6$\,fs).

\begin{figure}
\includegraphics[width=0.45\textwidth]{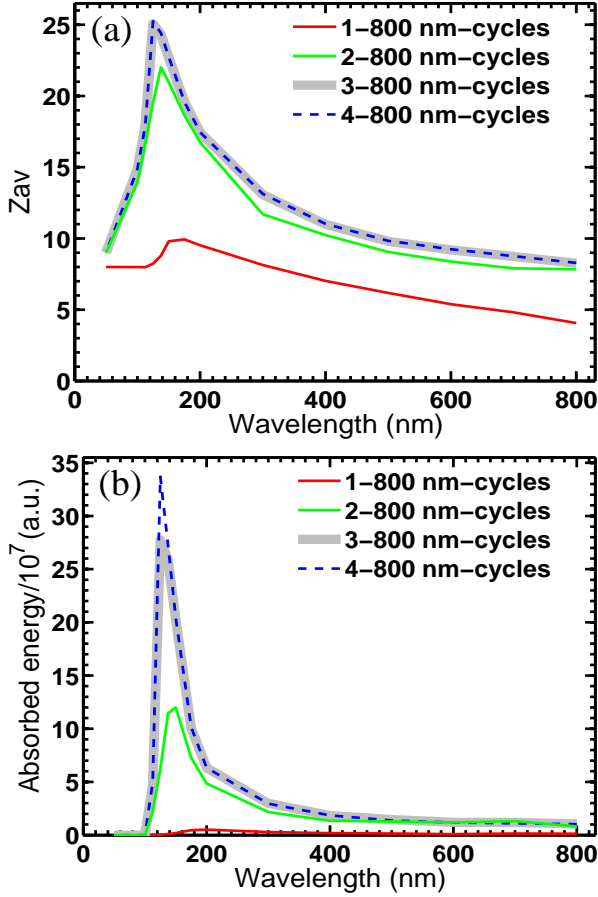}
\caption{(Color online)
Average ion charge $\chZav$ (a) and 
total absorbed energy $\Etot$ (b) vs laser wavelength after 
$1,2,3$ and $4$ laser cycles (at 800\,nm) for a 
$\mathrm{Xe_{17256}}$ cluster of radius $\Rinit \approx 7$~nm. 
Other parameters as in Fig.~\ref{Fig2Ch5}.
\label{Fig3Ch5}
}
\end{figure}

Figure~\ref{Fig3Ch5} shows the average charge state $\chZav$ and
total absorbed energy $\Etot$ vs the laser wavelength 
for the $\mathrm{Xe_{17256}}$ cluster of radius $\Rinit \approx 7$~nm 
after $t = 1, 2 , 3, 4$-laser cycles at 800\,nm. 
The value of $\chZav$ increases in time (in Fig.~\ref{Fig3Ch5}a) for all wavelengths. 
Ionization mostly occurs before $t=2$ cycles 
when the peak of the pulse is reached. After that the space charge field is high enough 
to generate further charge states between $2-3$ cycles.
$\chZav$ does not change anymore between $3-4$ cycles, indicating a 
saturation of inner ionization. The average charge state  $\chZav$ increases from $\chZ = 8$ to a maximum value $\chZav \approx 25$ 
as the laser wavelength is decreased from the infrared $800$~nm down to the UV wavelength $125$~nm.
It means that the sub-shells $4s^{2}p^{6}d^{10}5s^{2}p^{6}$ of almost all atoms are empty at $125$~nm.
A further decrease of the wavelength causes $\chZav$ to decrease gradually to a 
smaller value $\chZav \approx 9$ at $50$~nm. 

Figure~\ref{Fig3Ch5}b shows a similar qualitative behavior of the absorbed energy both in the time domain and in the 
wavelength domain. The energy $\Etot$ is maximum at the same wavelength $\lambdalaser = 125$~nm. Although the laser-pulse energy 
is the same in all cases the increased absorption at $125$~nm, leading to a marked 
increase of the average charge up to a value $\chZav \approx 25$ clearly 
shows that wavelength effects are undoubtedly important. One may compare the absorbed energy
and the average charge with the dual-pulse simulation 
results in Fig.~\ref{Fig2Ch5}. The absorbed energy $\Etot \approx 34\times10^{7}$\,a.u.\ and the 
average charge $\chZ\approx 25$ are 
much higher in the present case around the laser wavelength $125$~nm
compared to the respective values $\Etot\approx 9\times 10^{7}$ and $\chZav\approx 18.5$ in Fig.~\ref{Fig2Ch5}b.
The absorption is $\approx 3.78$ times higher than in Fig.~\ref{Fig2Ch5}b.
Moreover, in the dual-pulse case the  
total laser-pulse energy was twice higher. Therefore, the absorption efficiency is
augmented further by a factor of two.  

\begin{figure}
\includegraphics[width=0.45\textwidth]{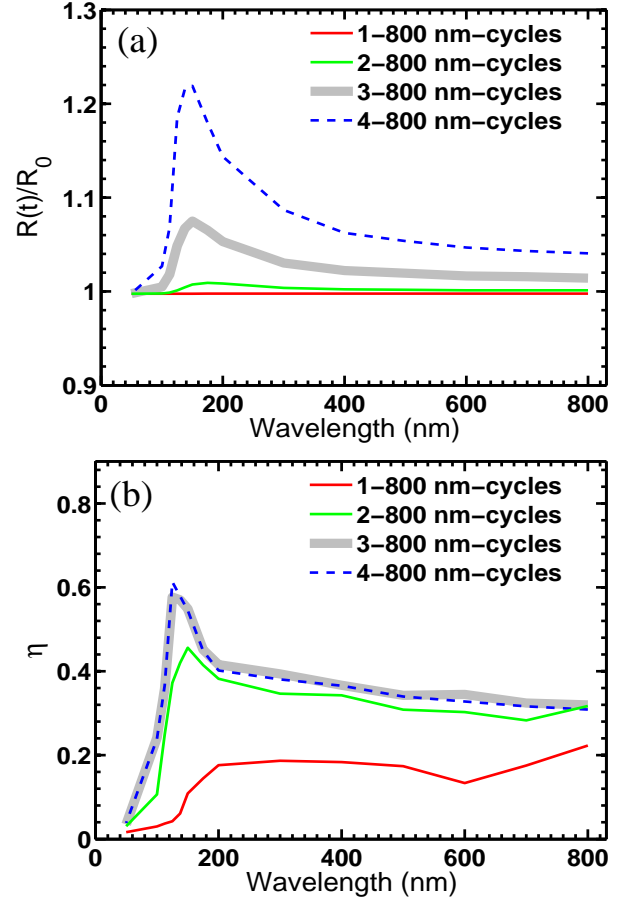}
\caption{(Color online)
Normalized cluster expansion radius $R/\Rinit$ (a) and 
outer ionization degree $\eta$~(b)
vs laser wavelength, corresponding 
to Fig.~\ref{Fig3Ch5}.
\label{Fig4Ch5}
}
\end{figure}

In Fig.~\ref{Fig4Ch5} we plot the normalized expansion radius $R(t)/\Rinit$ 
(Fig.~\ref{Fig4Ch5}a) 
and the outer ionization degree $\eta$ (number of total electrons outside 
$R(t)$ divided by the total number of electrons 
produced, $N\chZav(t)$, in Fig.~\ref{Fig4Ch5}b) 
vs the laser wavelength corresponding to the results 
in Fig.~\ref{Fig3Ch5}. The radius $R(t)$
and the outer ionization degree $\eta(t)$ go hand in hand with the 
absorbed energy $\Etot(t)$ and the charge $\chZav(t)$. 
After four cycles the cluster has expanded very little, $R(4T)/\Rinit \approx 1.225$ at $\lambdalaser \approx 125$~nm,
although the average charge $\chZav\approx 25$ is very high compared to Fig.~\ref{Fig2Ch5}. 
With such an insignificant 
expansion the space charge field can be considered optimized, leading to 
maximum ionization ignition. The 
ignition field (i.e., the space charge field due to the ionic background) under the assumption that   all electrons are removed reads
 $\Eig(t)\approx N\chZav(t)/R(t)^2$.
Using $R(t)$ from Fig.~\ref{Fig4Ch5}a and $\chZav(t)$ from Fig.~\ref{Fig3Ch5}a, 
one obtains at $125$~nm $\Eig(2T)\approx 20.0$, $\Eig(3T)\approx 21.0$ and 
$\Eig(4T)\approx 16.2$, if all electrons are removed 
(i.e., $\eta =1$). The expected ignition field $\approx 21$\,a.u.\ is maximum near 
the pulse peak around $2-3$-cycles, thereafter
decreases to $\Eig(4T)\approx 16.2$ due to an expansion 
$R(4T)/\Rinit \approx 1.225$ and no further creation of charge states.
Note that the peak laser field is only
$\Ehat\approx 1.19$. Therefore the enhanced ionization is certainly 
due to the ignition field. However, at $\lambdalaser\approx 125$~nm
$\eta\approx 0.6$ in Fig.~\ref{Fig4Ch5}b, meaning that $40\%$ of the electrons are still inside
the cluster.  The presence of these 
electrons lowers $\Eig$ compared to the above ideal case 
of $\eta = 100\%$ outer ionization, and one may argue that $\Eig$ is not yet optimized. 
However, even if $\eta = 100\%$ outer ionization is achieved for the above
laser field intensity the maximum total field is $\approx 22$ which is still 
insufficient to produce a higher average charge $\chZav=27$ 
(requiring a threshold field $\gtrapprox24$ according to OFI). Hence the average charge state is optimized. This will be shown explicitly  at the end of this section where we actually achieve  $\eta \approx 100\%$ for this cluster.

\begin{figure}
\includegraphics[width=0.485\textwidth]{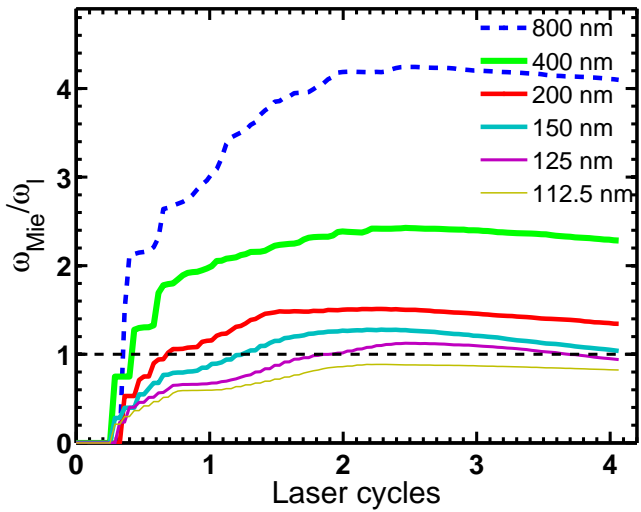}
\caption{(Color online)
Scaled Mie-frequency $\omegaMie/\omegalaser$
vs time (in 800\,nm cycles) for wavelengths $\lambdalaser = 800-112.5$~nm
and the laser and cluster parameters of Fig.~\ref{Fig3Ch5}. 
\label{Fig5Ch5}
}
\end{figure}

The above results clearly show that there exists a certain wavelength at which the laser-cluster coupling is very efficient. 
Such a nonlinear dependence of 
the absorbed energy and average charge state on the laser wavelength 
indicates a resonance around $125$~nm in Fig.~\ref{Fig3Ch5} and \ref{Fig4Ch5}. 
To investigate this further, we plot in Fig.~\ref{Fig5Ch5} the scaled Mie-frequency  $\omegaMie(t)/\omegalaser$ vs time. The dashed line indicates the linear resonance.
Charging of the cluster starts around $0.3$ cycles for all 
wavelengths by OFI, leading to an abrupt increase of $\omegaMie(t)/\omegalaser$ for the longer wavelengths while for the shorter ones the increase proceeds slower.
As a result the plasma is overdense during the entire pulse for the long wavelengths but stays close to the linear resonance for the shorter wavelengths. 
The more time is spent near the linear resonance, the higher is the energy 
absorption and the average charge state, as seen in Fig.~\ref{Fig3Ch5}.
At the wavelength $125$~nm the resonance is met at the peak of the pulse so that the energy absorption is particularly efficient.

\begin{figure}
\includegraphics[width=0.4\textwidth]{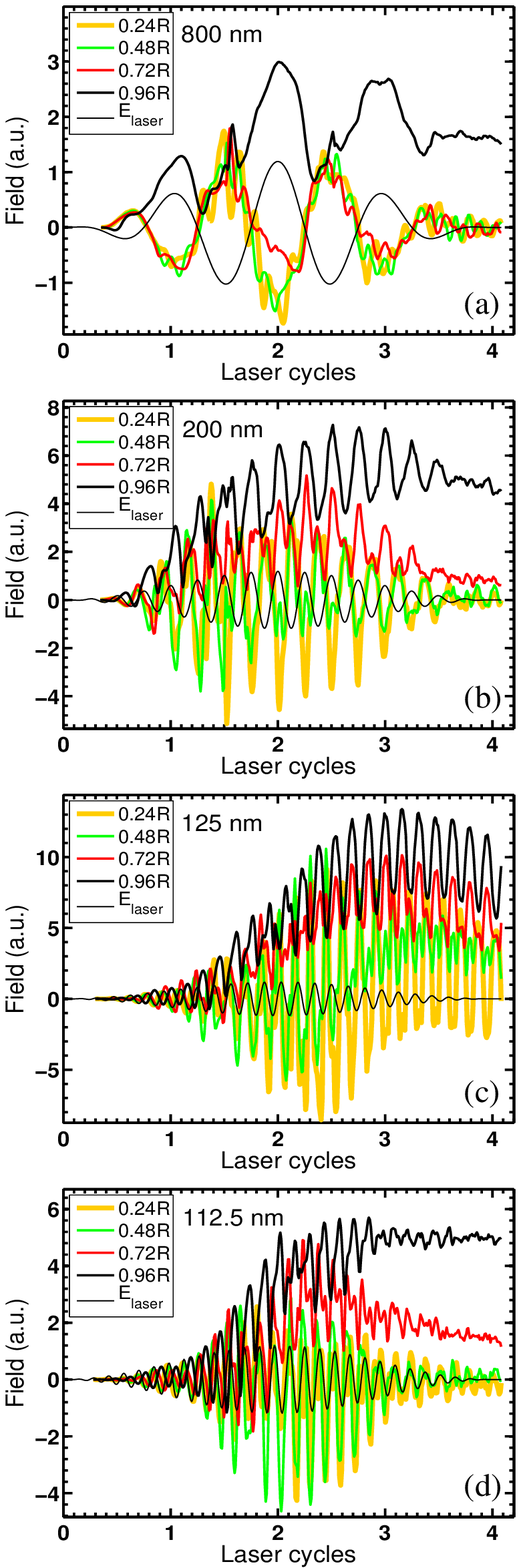}
\caption{(Color online)
The $x$-component of the space charge field 
at radial distances $0.24\Rinit, 0.48\Rinit, 0.72\Rinit$, and $0.96 \Rinit$ 
inside a $\mathrm{Xe_{17256}}$ cluster of radius $\Rinit \approx 7$~nm 
and the laser field
$\laserE(t)=\Ehat\sin^2(\pi t/nT)\cos(\omegalaser t)$ of  peak intensity  $5\times10^{16}\Wcmcm$
vs time (in periods corresponding to 800\,nm)  at (a) $\lambdalaser = 800$, 
(b) $\lambdalaser = 200$, 
(c) $\lambdalaser = 125$, and 
(d) $\lambdalaser = 112.5$~nm. 
\label{Fig6Ch5}
}
\end{figure}

\begin{figure}
\includegraphics[width=0.4\textwidth]{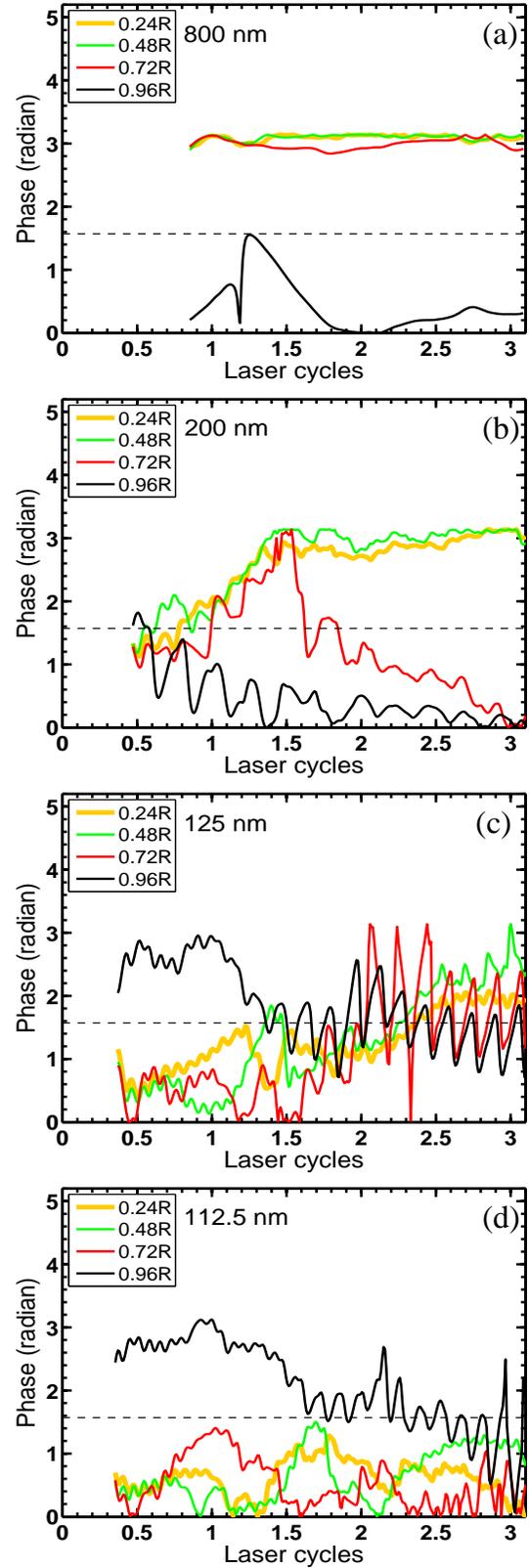}
\caption{(Color online)
Phase of the
space charge field with respect to the  laser field at different radial distances corresponding to Fig.~\ref{Fig6Ch5} vs time corresponding to Fig.~\ref{Fig6Ch5}.
\label{Fig7Ch5}
}
\end{figure}

We now discuss the time evolution of the space charge field $\Escx$ (along the laser polarization) at different positions
inside the cluster to further illustrate the  resonance at short wavelengths, leading to efficient ionization ignition and the generation of high charge states. Figure~\ref{Fig6Ch5} shows the space charge field $\Escx$ and the laser field $\laserE$ vs time
at radial positions $0.24\Rinit$, $0.48\Rinit$, $0.72\Rinit$ and $0.96\Rinit$ for four different wavelengths. Figure~\ref{Fig7Ch5} shows the corresponding phases of $\Escx$ with respect to the driving laser field.

At the long wavelength $800$~nm, $\Escx$  inside the cluster at radii
$0.24\Rinit$, $0.48\Rinit$, $0.72\Rinit$ mostly oscillates with a phase $\delta \approx \pi$ while $\delta \approx 0$  at the boundary (i.e., at $0.96\Rinit$).
This is clearly what one expects from an overdense plasma: screening of the laser field in the cluster interior but an opposite behavior outside the electron cloud. The oscillation of the space charge field arises 
due to the oscillations of the electrons trapped inside the cluster. These electrons form approximately a sphere which is smaller than the cluster due to outer ionization. If the electron cloud was rigid and did not cross the cluster boundary the phase should be exactly $\pi$ and $0$ inside and outside, respectively, if the plasma is overdense,  and opposite in the underdense case.  In reality, the bound electron population changes and the electron sphere is neither rigid nor has it a sharp boundary, resulting in  phase distortions and deviations  from the idealized case, as seen in Fig.~\ref{Fig7Ch5}a. 

Figure~\ref{Fig6Ch5} confirms explicitly that the total field at the boundary is highest and therefore leads to the highest ionic
charge states while $\Escx(t)$ almost nullifies the laser field in the strongly overdense regime.  
The maximum value of the total field 
at the peak of the pulse is $\approx 4.0$~a.u.\ ($x$-component only) which is sufficient to produce 
charge states up to $\chZ \approx 18$ (also seen in Fig.~\ref{Fig1Ch5}b). 
An additional 
contribution (up to a factor $\sqrt{3}$) to the total field comes from the $y$ and $z$-components of the space charge field.

At $200$~nm the amplitude of $\Escx$ around $t=1.5$~cycles at $0.24\Rinit$ increases up to $5$~a.u.\ which, after addition to 
the laser field, is sufficient to produce charge states $\chZ = 18$ even inside the cluster. 
After $\approx 1.75$~cycles $\Escx$ at $0.72\Rinit$ behaves similarly 
to that at $0.96\Rinit$, i.e., the laser and space charge fields at
$0.96\Rinit$ and $0.72\Rinit$ are now approximately in phase. Figure~\ref{Fig6Ch5}b shows that
the total field $\approx 4-8$ between $0.96\Rinit$ and $0.72\Rinit$, producing 
 charge states $\chZ \approx 18-23$. 
However, due to the screening of the laser field inside the cluster, many atoms  there have only charge states $\chZ <18$ so that the average charge state is $\chZav\approx 18$ in Fig.~\ref{Fig3Ch5}. 
From Fig.~\ref{Fig7Ch5}b it is seen that before $t=1.75$~cycles the phase of $\Escx$ at $0.72\Rinit$ approximately follows the phase at the smaller radii 
$0.24\Rinit$ and $0.48\Rinit$ since the plasma is evolving from under to overdense. Then, with increasing outer ionization and thus shrinking electron sphere, $\Escx$ at $0.72\Rinit$ drops and approaches the behavior for $0.96\Rinit$. 

At the resonant wavelength $125$~nm 
violent oscillations of the electron cloud are driven, leading to a particularly high total field everywhere inside the cluster and an average charge state  $\chZav \approx 25$ in Fig.~\ref{Fig3Ch5}.
Higher charge states $\chZ>26$ are not produced because of the high threshold field $\approx24$ necessary to crack the M-shell.
One may argue that the presence of $40\%$ electrons inside 
the cluster (in Fig.~\ref{Fig4Ch5}b)
will deplete the field inside significantly.
However, one should keep in mind that at resonance the electron cloud oscillates with a large excursion, 
exposing a substantial part of the naked ionic background, leading to an enhanced ``dynamical ionization ignition'' 
\cite{bauer03} 
which can produce higher charge states than expected from the laser field alone  even inside the cluster. 
Finally, after $t=3$~cycles $\Escx$ at the boundary drops due to the cluster expansion. As expected, the phases plotted in  Fig.~\ref{Fig7Ch5}c fluctuate around $\pi/2$ throughout the cluster once the resonance condition is met.

At $112.5$~nm the plasma remains underdense. Figure~\ref{Fig6Ch5}d shows that the space charge field amplitudes drop compared to those in Fig.~\ref{Fig6Ch5}c, yielding less ionization ignition and absorbed 
energy, similar to the $200$\,nm-case.

\begin{figure}
\includegraphics[width=0.45\textwidth]{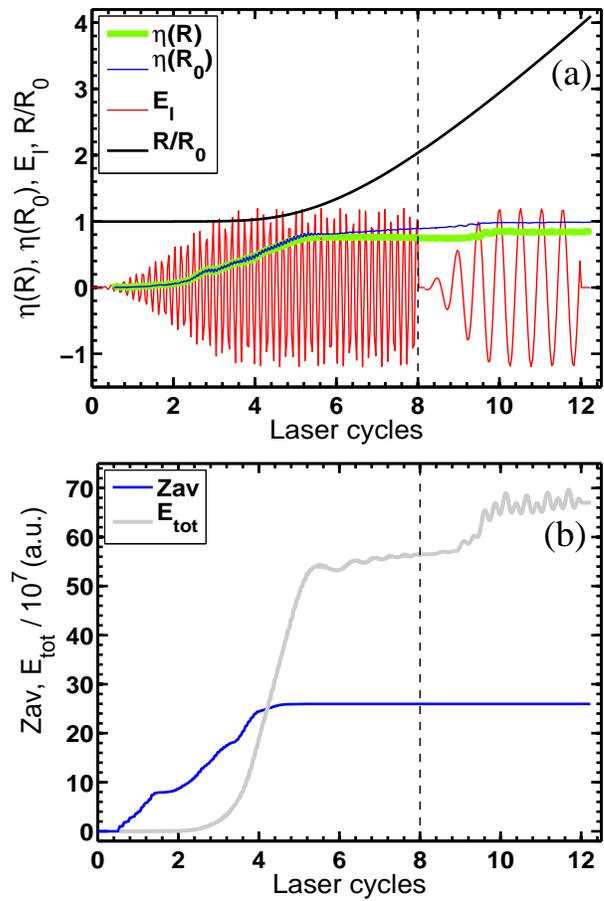}
\caption{(Color online)
Variation of (a)
normalized cluster radius $R/\Rinit$, 
degree of outer ionization $\eta(R), \eta(\Rinit)$ within $R$ and $\Rinit$, laser field
$\laserE$ (in atomic units) (b) average ion charge $\chZav$ and total absorbed energy $\Etot$ 
in time (in units of the laser period $T$ corresponding to $800$~nm)
for the $\mathrm{Xe_{17256}}$ cluster of Fig.~\ref{Fig3Ch5}.
The two laser pulses of wavelength 125 and 415\,nm, respectively, are included in (a). 
The peak intensity is the same as in Fig.~\ref{Fig3Ch5} for both pulses.
\label{Fig8Ch5}
}
\end{figure}

Even in the optimal $125$\,nm case presented so far only $60\%$ of the  generated electrons were removed from the cluster (visible in Fig.~\ref{Fig4Ch5}b).
Therefore outer ionization and ionization ignition
was certainly not optimized. We argued that even if the remaining $40\%$ electrons were removed, the average charge state would not be significantly increased as compared to that shown in Fig.~\ref{Fig3Ch5}a.
To prove that, we performed PIC simulations for the same cluster and the same peak intensity but now employing two consecutive pulses (shown in Fig.~\ref{Fig8Ch5}). The first pulse of resonant wavelength $125$\,nm with respect to the still compact cluster  is ramped up over four  $800\,$nm-cycles and held constant afterwards up to 8~cycles (the details of how the pulse is ramped down do not matter; therefore it is simply switched off abruptly). At $t=8$~cycles a second pulse is switched on (over 2~cycles) whose frequency is resonant with the Mie-frequency around $t=10$~cycles.

After the first pulse the cluster doubled its radius, and the outer ionization degree amounts to  $\eta(R)\approx 0.8$ so that
$20\%$ electrons are still inside the cluster of radius $R(t)$ while $\approx 10\%$ are inside a sphere of radius $\Rinit$. The average charge $\chZav$ in Fig.~\ref{Fig8Ch5}b does not change significantly compared to Fig.~\ref{Fig3Ch5}a although the pulse energy per unit area $\int_0^{8T}\laserE^2(t)\diff t$ is $\approx 3-4$ times higher.

The purpose of the second pulse shown in Fig.~\ref{Fig8Ch5}b is the removal of the residual electrons. Although almost $95\%$ outer ionization within the expanding radius $R$ and $99\%$ within $\Rinit$ are achieved, no higher charge states are created. The absorbed energy also does not rise significantly so that the higher input energy invested into the two pulses does not pay off. Hence a single, short UV-pulse of wavelength $125$~nm turns out to be optimal with respect to fractional energy absorption and generation of a high average charge state  under the conditions considered.

\section{Summary}\label{sec6Ch5}
In summary, we studied the interaction of xenon clusters with intense short 
laser pulses using a three-dimensional PIC code. Our aim was to 
optimize  for a given cluster the laser energy absorption and the generation of high average charge states. The latter will then lead to energetic ions upon Coulomb explosion.
We showed that for a given laser intensity an optimal laser wavelength exists that, under the typical conditions studied in this work, lies in the UV regime. Energy absorption is optimized  when resonance is met during an early stage of the dynamics when the cluster is still compact. The conventional, long-pulse linear resonance during the expansion of the cluster is less efficient.

\section*{Acknowledgments}
We thank Sergei Popruzhenko for valuable discussions and careful proofreading.
This work was supported by the Deutsche Forschungsgemeinschaft.

\end{document}